\documentclass[aps,prl,reprint,groupedaddress,showpacs]{revtex4-1}
\usepackage{graphicx}
\usepackage{dcolumn}
\usepackage{bm}
\usepackage[latin1]{inputenc}
\usepackage{color}
\newcommand{\vE}{\vec{\textbf{\emph{E}}}}

\begin{document}


\title{Optical response of a bilayer crystal}


\author{Luca Dell'Anna}
\author{Michele Merano}
\email[]{michele.merano@unipd.it}
\affiliation{Dipartimento di Fisica e Astronomia G. Galilei, Universit$\grave{a}$ degli studi di Padova, via Marzolo 8, 35131 Padova, Italy}


\date{\today}

\begin{abstract}
We extend the recently developed classical theory for the optical response of a single-layer crystal to bilayers. We account for the interaction between the two atomic planes and the multiple reflections inside the crystals. We show how to define a global susceptibility meaningful for the bilayer crystal and how its expression varies compared to the single-layer case. We compute both the local and the macroscopic fields, which allow us to make a direct comparison with experimental data. 
\end{abstract}


\maketitle
\section{Introduction}
A two-dimensional (2D) atomic crystal is a single plane of atoms or molecules whose properties are significantly different from those of its three-dimensional (3D) precursor. This is true from a thermodynamic point of view and it becomes impressive when we consider electronics properties. Charge carriers in graphene are massless Dirac fermions \cite{Novoselov2004}. A transition-metal dichalcogeneide monolayer is a direct band-gap semiconductor while bilayer, three-layer, and multi-layer crystals are indirect band-gap semiconductors \cite{Heinz2010}. These single-layer crystals have promise for a large number of applications \cite{Dhanabalan16} because they are stable under ambient conditions and, despite their reduced dimensionality, they are truly macroscopic objects \cite{Novoselov2005}.

In spite of their atomic layer thickness, these materials exhibit strong light-matter interaction \cite{Nair2008, Blake2007}. It was a surprising discovery that 2D crystal monolayers, deposited on suitable substrates, produce an optical contrast of up to several percent at specific wavelengths, making them easily visible \cite{Blake2007, Kis11}. It was some time before this phenomenon was fully comprehended and a proper theoretical description could be provided. The first analysis treated the single-layer crystal as a slab with an effective thickness \cite{Blake2007}. Only a few years later the adoption of the surface-current model allowed for a completely satisfactory analysis of the optical experiments on these crystals \cite{Pershoguba07, Hanson08, Zhan2013, Galina2014, Merano16}.   

The optical response of a single-layer crystal provides direct access to its electronic properties via its macroscopic surface susceptibility and surface conductivity \cite{Kravets2010, Heinz2014, Merano16, Merano316, Jayaswal18}. Recently a classical description of a 2D crystal connected these macroscopic quantities to microscopic atomic polarizability through the Clausius-Mossotti-Lorenz-Lorentz relations. First, a microscopic approach has shown that retardation effects are very relevant for the optical properties of these crystals \cite{Luca16}. Then, the computation of the macroscopic field has required the advanced potential solutions of the inhomogeneous Maxwell' s equations, via the radiation-reaction electric field \cite{Merano17}.

The first successful technique to produce two-dimensional materials was exfoliation \cite{Novoselov2005}. Now other growth methods are available, such as chemical vapor deposition \cite{Sutter08}. All these experimental techniques are able to produce 2D crystals with different numbers of layers starting from single-layer materials, to bilayers, three layers, and up to the bulk. Optical contrast experiments are able to distinguish between the number of constituent planes of a 2D crystal \cite{Blake2007, Blake2011}, but a proper theoretical analysis is still lacking for the bilayer case. 

In this paper, we aim to extend the complete classical physical picture that has been developed for the optical response of a monolayer crystal \cite{Luca16} to a bilayer material, i.e., two planes of atoms or molecules separated by a certain interlayer distance. In particular, we will address the following questions: How does the interaction between the two planes of a bilayer crystal influence its optical properties? How does this interaction scale with the distance between the atomic planes? Can we still use a surface susceptibility to describe a bilayer crystal, or do we need to introduce a volume susceptibility? We choose to treat bilayer hexagonal Boron Nitride because it is an insulating dielectric. From the standpoint of optics, this is the simplest example of a bilayer crystal.

\section{Classical model of radiating bilayer 2D crystals}

\begin{figure}
\includegraphics{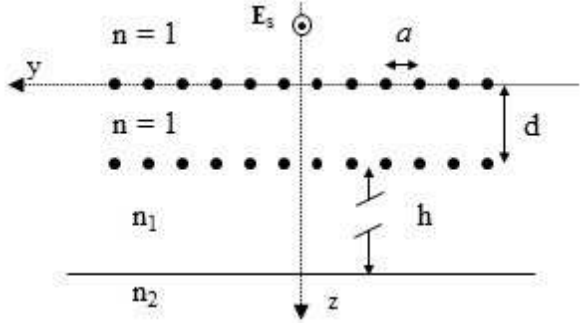}
\caption{A bilayer crystal is modelled as 2 two-dimensional Bravais lattices in vacuum (n=1). A linearly polaraized plane wave is incident on it from vacuum. The crystal can be free-standing, deposited on a bulk substrate $n_1=n_2$, or on a stratified substrate $n_1\neq n_2$. $a$ denotes the lattice spacing, $d$ is the interlayer distance, and $h$ is the thickness of the medium with refractive index $n_1$.}
\label{fig1}
\end{figure}
We consider an insulating free-standing bilayer 2D crystal formed by $N$ atoms per unit area placed on two 2D Bravais lattices with lattice spacing $a$ (Fig. \ref{fig1}). The two atomic planes are separated by a distance $d$ and each atom has a polarizability $\alpha$. A linearly polarized (along the $x$-direction) electromagnetic plane wave is incident on this 2D crystal with an harmonic time dependence $e^{i\omega t}$. For the sake of simplicity, normal incidence is assumed and the crystal is initially supposed to be in the vacuum ($n_1=n_2=1$). As a consquence of electromagnetic excitation, the atoms of the crystal act as oscillating dipoles,
\begin{equation}
\label{Local3}
\vec{\textbf{\emph{p}}}_1(t)=\alpha \epsilon_{0} \vec{\textbf{\emph{E}}}^{(1)}_{loc}e^{i\omega t}\,,\;
\quad \vec{\textbf{\emph{p}}}_2(t)=\alpha \epsilon_{0} \vec{\textbf{\emph{E}}}^{(2)}_{loc}e^{i\omega t}
\end{equation}
where $\epsilon_0$ is the vacuum permittivity, and $\vec{\textbf{\emph{p}}}_1$, $\vec{\textbf{\emph{p}}}_2$, and $\vec{\textbf{\emph{E}}}^{(1)}_{loc}$, $\vec{\textbf{\emph{E}}}^{(2)}_{loc}$ are respectively the induced dipole moments and the local fields in the first and the second layer. The superposition principle provides $\vec{\textbf{\emph{E}}}^{(1)}_{loc}$ and $\vec{\textbf{\emph{E}}}^{(2)}_{loc}$
\begin{eqnarray}
\label{Local}
\vec{\textbf{\emph{E}}}^{(1)}_{loc}e^{i\omega t}&=&\vec{\textbf{\emph{E}}}_{i}e^{i\omega t}
+{\sum_{(m,n)}}'\vec{\textbf{\emph{E}}}^{(1)}_{n,m}(t)+{\sum_{(m,n)}}''\vec{\textbf{\emph{E}}}^{(2)}_{n,m}(t) \\
\vec{\textbf{\emph{E}}}^{(2)}_{loc}e^{i\omega t}&=&\vec{\textbf{\emph{E}}}_{i}e^{i\omega (t-d/c)}
+{\sum_{(m,n)}}'\vec{\textbf{\emph{E}}}^{(2)}_{n,m}(t)+{\sum_{(m,n)}}''\vec{\textbf{\emph{E}}}^{(1)}_{n,m}(t) \nonumber
\end{eqnarray}
where $\vec{\textbf{\emph{E}}}_{i}$ is the incident electric field and the sums $\sum '$ count the contributions coming from all the other dipoles of the same layer, while $\sum''$ count those coming from all the dipoles of the other layer. 
The expression of the dipole fields is  
\begin{equation}
\label{Dipole1}
\vec{\textbf{\emph{E}}}^{(i)}_{n,m}(
t)=\frac{1}{4\pi \epsilon_{0}r^3} \biggl(3(\tilde{\vec{\textbf{\emph{p}}}}_i\cdot\hat{\textbf{\emph{r}}})\hat{\textbf{\emph{r}}}-
\tilde{\vec{\textbf{\emph{p}}}}_i -\frac{( \vec{\textbf{\emph{r}}} \times \ddot{\vec{\textbf{\emph{p}}}}_i)  \times \vec{\textbf{\emph{r}}}}{c^2}  \biggr)
\end{equation}
where  
\begin{eqnarray}
\vec{\textbf{\emph{p}}}_i&=&\vec{\textbf{\emph{p}}}_i(t - \frac{r}{c})=\vec{\textbf{\emph{p}}}_ie^{i(\omega t -kr)} \nonumber \\
\tilde{\vec{\textbf{\emph{p}}}}_i&=&\vec{\textbf{\emph{p}}}_i(t - \frac{r}{c}) + \frac{r}{c}\dot{\vec{\textbf{\emph{p}}}}_i(t - \frac{r}{c})=\vec{\textbf{\emph{p}}}_ie^{i(\omega t -kr)}(1+ikr) \nonumber
\end{eqnarray}
with $i=1,2$ and where $(n,m)$ label the lattice sites located at $\textbf{\emph{r}}\equiv \textbf{\emph{r}}_{n,m}$.

We will first develop a microscopic theory to compute the local fields $\vec{\textbf{\emph{E}}}^{(1)}_{loc}$, $\vec{\textbf{\emph{E}}}^{(2)}_{loc}$. Afterwards, we will consider the macroscopic theory to obtain first the polarization,
\begin{equation}
\label{Phoney}
\vec{\textbf{\emph{P}}}=\frac{\vec{\textbf{\emph{P}}}_1+\vec{\textbf{\emph{P}}}_2}{d}=\frac{N}{d}(\vec{\textbf{\emph{p}}}_1+\vec{\textbf{\emph{p}}}_2)
\end{equation}
and then the macroscopic electric field $\vec{\textbf{\emph{E}}}$ and the electric susceptibility $\chi$ through
\begin{equation}
\label{Phoney2}
\vec{\textbf{\emph{P}}}=\epsilon_0\chi\vec{\textbf{\emph{E}}}.
\end{equation} 

\section{microscopic theory} 
We perform the sums in Eqs.~(\ref{Local}) by dividing the components parallel to the crystal planes from those perpendicular to them
\begin{eqnarray}
\nonumber \sum_{(n,m)}'\vec{\textbf{\emph{E}}}^{(i)}_\parallel
&=&\frac{\alpha}{4\pi}\vE_{loc\parallel}^{(i)}\\
&&\hspace{-1.3cm}\sum_{(m,n)\neq (0,0)}\hspace{-0.1cm}\left\{e^{-ikr_{nm}}
\left(
\frac{1+ik\,r_{nm}+k^2r^2_{nm}}{2\,r^{3}_{nm}}\right)\right\}
\label{sum1}
\\
\nonumber\sum_{(n,m)}'\vec{\textbf{\emph{E}}}^{(i)}_\perp
&=&-\frac{\alpha}{4\pi}\vE_{loc\perp}^{(i)}\\
&&\hspace{-1.3cm}\sum_{(m,n)\neq (0,0)}\hspace{-0.1cm}\left\{e^{-ikr_{nm}}
\left(
\frac{1+ik\,r_{nm}-k^2r^2_{nm}}{r^{3}_{nm}}\right)\right\}
\label{sum1perp}
\\
\nonumber \sum_{(n,m)}''\vec{\textbf{\emph{E}}}^{(i)}_\parallel
&=&\frac{\alpha}{4\pi}\vE_{loc\parallel}^{(i)}
\\&&\nonumber
\hspace{-1.cm}\sum_{(m,n)}\left\{\frac{e^{-ikr'_{nm}}}
{2\,r'^{3}_{nm}}
\Big[\left(
1+ik\,r'_{nm}+k^2r'^2_{nm}
\right)\right.\\
&&\left.
-\frac{3\,d^2}{r'^2_{nm}}(1+ik\,r'_{nm})+{d^2k^2}\Big]
\right\}
\label{sum2}
\\
\nonumber \sum_{(n,m)}''\vec{\textbf{\emph{E}}}^{(i)}_\perp
&=&-\frac{\alpha}{4\pi}\vE_{loc\perp}^{(i)}
\\&&\nonumber
\hspace{-1.cm}\sum_{(m,n)}\left\{\frac{e^{-ikr'_{nm}}}
{r'^{3}_{nm}}
\Big[\left(
1+ik\,r'_{nm}-k^2r'^2_{nm}
\right)\right.\\
&&\left.
-\frac{3\,d^2}{r'^2_{nm}}(1+ik\,r'_{nm})-{d^2k^2}\Big]
\right\}.
\label{sum2perp}
\end{eqnarray}
One can easily find that ${\vE_{loc\perp}^{(i)}}=0$ for a normally incident electromagnetic wave.

\subsection{Square and triangular lattice}
For the parallalel components, we obtain numerical results consistent with the following expressions for the local fields
\begin{eqnarray}
\label{Bilayer_Local1}
E^{(1)}_{loc}&=&E_{i}+\frac{\alpha}{4\pi a^3}\bigg[\big(C_0+i\, C_1 k a\big)
E^{(1)}_{loc}+ \nonumber \\ 
&&+\big(C_{d}+i\, C_1 k a e^{-ikd}\big)E^{(2)}_{loc}\bigg]\\
E^{(2)}_{loc}&=&E_{i}e^{-ikd}+\frac{\alpha}{4\pi a^3}\bigg[\big(C_0+i\, C_1 
k a\big)E^{(2)}_{loc}+ \nonumber \\ 
&&+\big(C_{d}+i\, C_1 k a e^{-ikd}\big)E^{(1)}_{loc}\bigg]
\label{Bilayer_Local2}
\end{eqnarray}
where the terms proportional to $E^{(1)}_{loc}$ in eq. (\ref{Bilayer_Local1}) and to $E^{(2)}_{loc}$ in eq. (\ref{Bilayer_Local2}) come from the sums $\sum '$ in eq. (\ref{Local}) and have already been computed in ref. \cite{Luca16}. The terms proportional to $E^{(2)}_{loc}$ in eq. (\ref{Bilayer_Local1}) and to $E^{(1)}_{loc}$ in eq. (\ref{Bilayer_Local2}) come from the sums $\sum ''$ in eq. (\ref{Local}). 

For both the square and the triangular lattice we find that $C_1=-2\pi N a^2$. For the square lattice we have
\begin{eqnarray}
C_0 &=& \hspace{-0.2cm}\sum_{(m,n)\neq(0,0)} \frac{1}{2(n^2+m^2)^\frac{3}{2}} \approx 4.517 \nonumber \\ 
C_d &=& {\hspace{0.2cm}\sum_{(m,n)}
\frac{(n^2+m^2-2\,d^2/a^2)}{2(n^2+m^2+d^2/a^2)^{\frac{5}{2}}}}
\end{eqnarray} 
where $C_0=2\zeta(3/2)\beta(3/2)$, with $\zeta(s)=\sum_{n=1}^\infty 1/n^s$ the Riemann zeta function and $\beta(s)=\sum_{n=0}^\infty (-1)^n/(2n+1)^s$ the Dirichlet beta function \cite{Luca16}. 
 For the triangular lattice 
\begin{eqnarray}
\nonumber C_0&=&\hspace{-0.2cm}\sum_{(m,n)\neq(0,0)}\frac{1}{2(n^2+nm+m^2)^{\frac{3}{2}}}\approx 5.517\\
C_d&=&\hspace{0.2cm}\sum_{(m,n)}
\frac{(n^2+nm+m^2-2\,d^2/a^2)}{2(n^2+nm+m^2+d^2/a^2)^{\frac{5}{2}}}
\end{eqnarray} 
Also in this case $C_0$ can be written in terms of special functions, $C_0=3\zeta(3/2)L(3/2,\chi_3)$, with $L(s,\chi_n)$ are Dirichlet L-series \cite{Luca16}.
\begin{figure}
\includegraphics{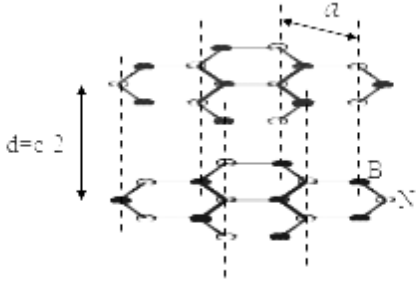}
\caption{Atomic structure of hexagonal Boron Nitride. The lattice paramters are $a\approx 0.25\,nm$ and $c\approx 0.67\,nm$ \cite{Pease52}.}
\label{Structure}
\end{figure}

\subsection{Honeycomb lattice}
Let us consider a special case of bipartite lattice, where there are two different atoms in the unit cell for the single monolayers \cite{Pease52}. In this case we have to generalize Eqs.~(\ref{Local}) for the local fields in the bilayer, introducing four terms, $\vec{\textbf{\emph{E}}}^{(i,j)}_{loc}$, where $i=1,2$ denotes the layers and $j=1,2$ labels the species of atoms with two different polarizabilities $\alpha_1$ and $\alpha_2$. For the structure depicted in Fig.~\ref{Structure}, as in the case of hexagonal boron nitride, 
Eqs.~(\ref{Bilayer_Local1}),~(\ref{Bilayer_Local2}) should be modified as it follows
\begin{eqnarray}
\label{hexagon1}
E^{(1,1)}_{loc}&=&E_{i}+\frac{1}{4\pi a^3}\bigg[\alpha_1
\big(C^{(1)}_{0}+i\, C_1 k a\big)E^{(1,1)}_{loc}+ \nonumber \\
\nonumber&&+\alpha_2\big(C^{(2)}_{0}+i\, C_1 k a \big)E^{(1,2)}_{loc}\\
\nonumber&&+\alpha_1\big(C^{(2)}_{d}+i\, C_1 k a e^{-ikd}\big)E^{(2,1)}_{loc}\\
&&+\alpha_2\big(C^{(1)}_{d}+i\, C_1 k a e^{-ikd}\big)E^{(2,2)}_{loc}\bigg]\\
\label{hexagon2} E^{(1,2)}_{loc}&=&E_{i}+\frac{1}{4\pi a^3}\bigg[\alpha_2
\big(C^{(1)}_{0}+i\, C_1 k a\big)E^{(1,2)}_{loc}+ \nonumber \\
\nonumber&&+\alpha_1\big(C^{(2)}_{0}+i\, C_1 k a \big)E^{(1,1)}_{loc}\\
\nonumber&&+\alpha_2\big(C^{(2)}_{d}+i\, C_1 k a e^{-ikd}\big)E^{(2,2)}_{loc}\\
&&+\alpha_1\big(C^{(1)}_{d}+i\, C_1 k a e^{-ikd}\big)E^{(2,1)}_{loc}\bigg]\\
\label{hexagon3} E^{(2,1)}_{loc}&=&E_{i}e^{-ikd}+\frac{1}{4\pi a^3}\bigg[\alpha_1
\big(C^{(1)}_{0}+i\, C_1 k a\big)E^{(2,1)}_{loc}+ \nonumber \\
\nonumber&&+\alpha_2\big(C^{(2)}_{0}+i\, C_1 k a \big)E^{(2,2)}_{loc}\\
\nonumber&&+\alpha_1\big(C^{(2)}_{d}+i\, C_1 k a e^{-ikd}\big)E^{(1,1)}_{loc}\\
&&+\alpha_2\big(C^{(1)}_{d}+i\, C_1 k a e^{-ikd}\big)E^{(1,2)}_{loc}\bigg]\\
E^{(2,2)}_{loc}&=&E_{i}e^{-ikd}+\frac{1}{4\pi a^3}\bigg[\alpha_2
\big(C^{(1)}_{0}+i\, C_1 k a\big)E^{(2,2)}_{loc}+ \nonumber \\
\nonumber&&+\alpha_1\big(C^{(2)}_{0}+i\, C_1 k a \big)E^{(2,1)}_{loc}\\
\nonumber&&+\alpha_2\big(C^{(2)}_{d}+i\, C_1 k a e^{-ikd}\big)E^{(1,2)}_{loc}\\
&&+\alpha_1\big(C^{(1)}_{d}+i\, C_1 k a e^{-ikd}\big)E^{(1,1)}_{loc}\bigg]
\label{hexagon4}
\end{eqnarray}
For $\alpha_1=\alpha_2$, we can identify $E^{(1,1)}$ with $E^{(1,2)}$ and $E^{(2,1)}$ with $E^{(2,2)}$, reducing to Eqs.~(\ref{Bilayer_Local1}),~(\ref{Bilayer_Local2}). The coefficients are $C_1=-2\pi N a^2$ and 
\begin{eqnarray}
\nonumber C^{(1)}_0&=&\hspace{-0.2cm}\sum_{(m,n)\neq(0,0)}\frac{1}{2(n^2+nm+m^2)^{\frac{3}{2}}}\approx 5.517\\
\nonumber C^{(2)}_0&=&\sum_{(m,n)}\frac{1}{2(n^2+nm+m^2+n+\frac{1}{3})^{\frac{3}{2}}}\approx 11.575\\
\nonumber C^{(1)}_d&=&\hspace{0.2cm}\sum_{(m,n)}
\frac{(n^2+nm+m^2-2\,d^2/a^2)}
{2(n^2+nm+m^2+d^2/a^2)^{\frac{5}{2}}}\\
\nonumber C^{(2)}_d&=&\hspace{0.2cm}\sum_{(m,n)}
\frac{(n^2+nm+m^2+n+1/3-2\,d^2/a^2)}
{2(n^2+nm+m^2+n+1/3+d^2/a^2)^{\frac{5}{2}}}
\end{eqnarray}
For $d/a= 4/3$, as in the case of the hexagonal boron nitride, we get 
$C^{(1)}_d\approx -0.010$ and $C^{(2)}_d\approx 0.005$.  Notice that $C_0^{(1)}$, $C_d^{(1)}$ and $C_1$ are the same as those of the triangular lattice. 

\subsection{Dependence of the interaction of the atomic planes on the distance $d$} 
\begin{figure}
\includegraphics[width=0.5\textwidth]{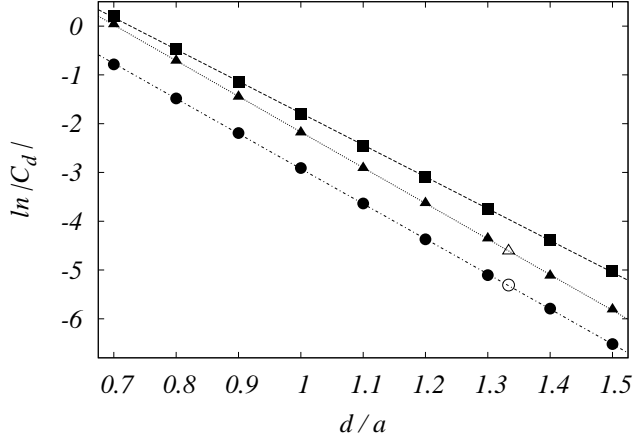}
\caption{$C_d$ (absolute value) for square lattice (squares), $C_d=C_d^{(1)}$ for triangular and honeycomb 
lattice (triangles) and $C_d^{(2)}$ for honeycomb lattice (circles), in logarithmic scale, as functions of the distance $d$ in units of $a$, the lattice parameter. The lines are obtained by fitting the numerical values by Eq.~(\ref{Cd}), where $A_\ell\approx -114.80$, $\lambda_\ell\approx a/6.53$ for the 
square lattice (dashed line), $A_\ell\approx -170.75$, $\lambda_\ell\approx a/7.31$ for the triangular and honeycomb lattices (dotted line), and $A_\ell\approx 71.38$, $\lambda_\ell\approx a/7.19$ for $C_d^{(2)}$ appearing in the honeycomb lattice (dotted-dashed line). The empty points are the values of $\ln |C_d^{(1)|}$ (triangle) and $\ln|C_d^{(2)}|$ (circle) for the bilayer hBN.}
\label{fig.Cd}
\end{figure}

\begin{figure}
\includegraphics[width=0.5\textwidth]{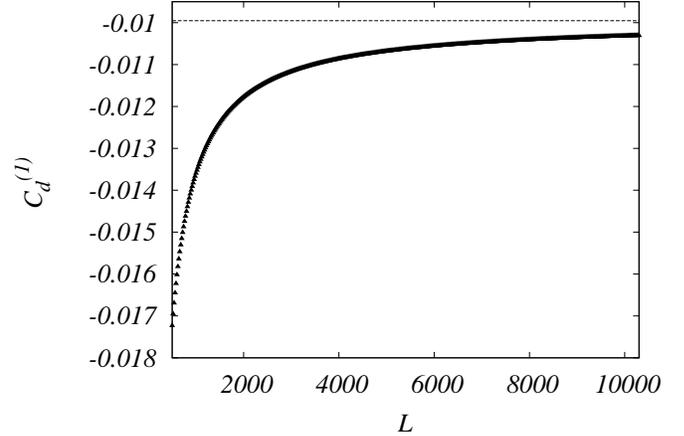}
\caption{Convergence for $C_d^{(1)}$ at $d/a=4/3$ (as in the case of hBN) obtained summing over $2L\times 2L$ sites ($-L \le n,m \le L$). The asimptotic value is obtained by fitting the points with the function $a+b/L$, getting $a=-0.00995$ (the asymptotic value for $C_d^{(1)}$, dotted line) and $b=-3.64$.}
\label{fig.convergence}
\end{figure}

Apart from the phase factor $e^{-ikd}$ that is due to the propagation of the electromagnetic radiation in vacuum, the only term in eqs. (\ref{Bilayer_Local1}) and (\ref{Bilayer_Local2}) (or in eqs. (\ref{hexagon1})-(\ref{hexagon4})) that depends on the distance $d$ between the two atomic planes is $C_d$ ($C_d^{(1)}$, $C_d^{(2)}$). We interpret it as a coefficient describing the interaction between the two atomic planes. Its dependence on the distance $d$ (in units of $a$) is shown in Fig.~\ref{fig.Cd}. In all the cases, the form of $C_d$, for large enough $d$ (Fig. \ref{fig.Cd}), fits well with the expression
\begin{equation}
\label{Cd}
C_d\simeq A_{\ell}\, \exp\left(-\frac{d}{\lambda_\ell}\right)
\end{equation}
where $A_{\ell}$ and $\lambda_\ell$ depend on the lattice, $A_{\ell}<0$ for the square lattice and the triangular lattice (for $C_d^{(1)}$), while it is $A_\ell>0$ in the case of $C_d^{(2)}$ in the honeycomb lattice (see Fig.~\ref{fig.Cd} where the values of $A_{\ell}$ and $\lambda_\ell$ in the three cases are reported). From eq. (\ref{Cd}) and Fig.~\ref{fig.Cd}, one can see that, as soon as $d$ far exceeds $a$, $C_d$ becomes negligible. The points in Fig.~\ref{fig.Cd} are obtained by finite size scaling as shown in Fig.~\ref{fig.convergence}, where the convergence of the sum for $C_d^{(1)}$ at $d\approx 1.333$ (useful for hBN) is reported as an example. 

\section{macroscopic theory}
We proceed from the local fields to the macroscopic fields by using an approach similar to the one developed in \cite{Merano17, Merano18}. The macroscopic fields to be computed are the reflected and the transmitted fields $E_{r}$, $E_{t}$, plus the resultant of all the positive (negative) traveling waves between the two planes $E_{+}$ ($E_{-}$). 
These fields must obey the boundary conditions. 

\subsection{Square and triangular lattice}
\subsubsection{Non-interacting case}
We note that the macroscopic surface current on the first (second) atomic plane is given by  \cite{Merano17, Merano18} $i\frac{k}{\eta}N\alpha E^{(i)}_{loc}=i\omega P_i $ ($i=1,2$). For the sake of simplicity we first consider the case of a distance $d$ in between the two atomic planes big enough to have $C_d$=0. The boundary conditions link together the macroscopic and the microscopic fields.
\begin{eqnarray}
\label{Bilayer_non-interacting}
&E&_{i}+E_{r}=E_{+}+E_{-} \\ &E&_{i}+E_{r}=E^{(1)}_{loc}(1-\frac{\alpha C_{0}}{4\pi a^3}) \nonumber  \\
&H&_{i}-H_{r}=H_{+}-H_{-}+i\frac{k}{\eta}N\alpha E^{(1)}_{loc}=H_{+}-H_{-}+i\omega P_1 \nonumber \\
&E&_{+}e^{-ikd}+E_{-}e^{ikd}=E_{t} \nonumber \\ &E&_{t}=E^{(2)}_{loc}(1-\frac{\alpha C_{0}}{4\pi a^3}) \nonumber \\
&H&_{+}e^{-ikd}-H_{-}e^{ikd}=H_{t}+i\frac{k}{\eta}N\alpha E^{(2)}_{loc}=H_{t}+i\omega P_2 \nonumber
\end{eqnarray}
The relation between $\vec{\emph{\textbf{E}}}$ and $\vec{\emph{\textbf{H}}}$ is $\eta\vec{\textbf{\emph{H}}}=\hat{s}\wedge\vec{\textbf{\emph{E}}}$, $\hat{s}$ is the unit vector along the propagation direction and $\eta$ is the impedance of vacuum. We have 6 equations and 6 unknown variables: $E_{r}$, $E_{t}$, $E_{+}$, $E_{-}$, $E^{(1)}_{loc}$, and $E^{(2)}_{loc}$. This approach is self-consistent becuase the solutions for $E^{(1)}_{loc}$ and $E^{(2)}_{loc}$ are identical to those provided by the microscopic equations (\ref{Bilayer_Local1}), (\ref{Bilayer_Local2}) with $C_{d}=0$. The macroscopic field in the first layer is naturally given by $E_{i}+E_{r}$ and in the second layer by $E_{t}$. We note that for both layers the local field is given by the macroscopic field divided by $(1-\frac{\alpha C_{0}}{4\pi a^3})$. In this sense it has the same expression of the monolayer \cite{Luca16}.

\subsubsection{Interacting case}
In the case of interacting atomic planes ($C_d \neq$ 0), the system of eqs. (\ref{Bilayer_non-interacting}) changes because (eqs. (\ref{Bilayer_Local1}), (\ref{Bilayer_Local2})) the field applied to one plane induces a local field and hence a polarization also on the other plane.
\begin{eqnarray}
\label{Bilayer_interacting}
&E&_{i}+E_{r}=E_{+}+E_{-} \\ &E&_{i}+E_{r}=E^{(1)}_{loc}(1-\frac{\alpha C_{0}}{4\pi a^3})-E^{(2)}_{loc}\frac{\alpha C_{d}}{4\pi a^3}  \nonumber \\
&H&_{i}-H_{r}=H_{+}-H_{-}+i\frac{k}{\eta}N\alpha E^{(1)}_{loc}=H_{+}-H_{-}+i\omega P_1  \nonumber \\
&E&_{+}e^{-ikd}+E_{-}e^{ikd}=E_{t};\nonumber \\ &E&_{t}=E^{(2)}_{loc}(1-\frac{\alpha C_{0}}{4\pi a^3})-E^{(1)}_{loc}\frac{\alpha C_{d}}{4\pi a^3} \nonumber \\
&H&_{+}e^{-ikd}-H_{-}e^{ikd}=H_{t}+i\frac{k}{\eta}N\alpha E^{(2)}_{loc}=H_{t}+i\omega P_2 \nonumber
\end{eqnarray}
Also for the interacting case self-consistency with the microscopic equations (\ref{Bilayer_Local1}), (\ref{Bilayer_Local2}) is verified. Importantly, now the local fields are coupled to the macroscopic fields of both layers. Their expression is now different from the one found for the monolayer. For $C_d=$ 0 we recover the non-interacting case. 

\subsection{Honeycomb lattice}
We note that the macroscopic surface current on the first (second) atomic plane is given by $i\frac{k}{\eta}N(\alpha_1 E^{(i, 1)}_{loc}+\alpha_2 E^{(i, 2)}_{loc})=i\omega P_i$ ($i=1,2$) \cite{Merano17}. In this case we have 8 equations and 8 unknown variables: $E_{r}$, $E_{t}$, $E_{+}$, $E_{-}$, $E^{(1, 1)}_{loc}$, $E^{(1, 2)}_{loc}$, $E^{(2, 1)}_{loc}$, and $E^{(2, 2)}_{loc}$. These equations are reported in Appendix I. The solutions for the local fields are self-consistent with the solutions of eqs. (\ref{hexagon1})-(\ref{hexagon4})

\section{The susceptibility of an insulating bilayer crystal}
From eq. (\ref{Phoney2}) we have
\begin{equation}
\label{chi}
\chi=\frac{ P_{1x}+ P_{2x}}{d\epsilon_0(E_{i}+E_{r}+E_{t})}.
\end{equation}
For the square and the triangular lattices,
\begin{equation}
\label{chi_square}
\chi=\frac{4\pi a^3N\alpha}{4\pi a^3d-(C_0+C_{d})d\alpha}.
\end{equation}
For the honeycomb lattice,
\begin{equation}
\label{chihoney}
\chi=\frac{N\Big(\frac{\alpha_1+\alpha_2}{d}-\frac{\alpha_1\alpha_2\left(C_0^{(1)}-C_0^{(2)}-C_d^{(1)}+C_d^{(2)}\right)}{2\pi a^3d}\Big)}{1-\frac{(C_0^{(1)}+C_d^{(2)})(\alpha_1+\alpha_2)}{4\pi a^3}+\frac{\alpha_1\alpha_2\left((C_0^{(1)}+C_d^{(2)})^2-(C_0^{(2)}+C_d^{(1)})^2\right)}{(4\pi a^3)^2}}
\end{equation}
Our calculations indicate that in the interacting case it is no more meaningful to provide a $\chi_{s}$ for each single plane (we would obtain different results for the two planes), but only a global $\chi$. For the non-interacting case this is of course still possible
\begin{equation}
\label{chis}
\chi_s=\frac{ P_{1x}}{\epsilon_0(E_{i}+E_{r})}=\frac{P_{2x}}{\epsilon_0 E_{t}}=\frac{ P_{1x}+ P_{2x}}{\epsilon_0(E_{i}+E_{r}+E_{t})}=\chi \, d
\end{equation}
For the square and triangular lattices,
\begin{equation}
\label{chis_square}
\chi_s =\frac{4\pi a^3N\alpha}{4\pi a^3-C_0\alpha}
\end{equation}
while for the honeycomb lattice,
\begin{equation}
\label{chis_honey}
\chi_s=\frac{N\Big(\alpha_1+\alpha_2-\frac{\alpha_1\alpha_2\left(C_0^{(1)}-C_0^{(2)}\right)}{2\pi a^3}\Big)}{1-\frac{C_0^{(1)}(\alpha_1+\alpha_2)}{4\pi a^3}+\frac{\alpha_1\alpha_2\left(C_0^{(1)2}-C_0^{(2)2}\right)}{(4\pi a^3)^2}}
\end{equation}
These last two expressions are of course equal to the surface susceptibilities of the monolayers. Looking at the expressions (\ref{chi_square}) and (\ref{chis_square}) we can compare the surface susceptibility of a monolayer with the susceptibility of the bilayer, and we have
\begin{equation}
\label{chivschis}
\frac{\chi_{s}}{d}>\chi
\end{equation}
The same relation holds for the hexagonal BN with reasonable assumptions for the atomic polarizabilities (see below). For the special case of square lattices ($N=1/a^2$) and $d=a$, we obtain
\begin{equation}
\label{chi_bilayer_sq}
\chi=\frac{\mathcal{N}\alpha}{1-\frac{(C_0+C_{d})\mathcal{N}\alpha}{4\pi}}
\end{equation}
where $\mathcal{N}=1/a^3$. As expected, $\chi$ is closer than $\chi_s /a$ to the susceptivity of the bulk $\chi_{3D}=\mathcal{N}\alpha/(1-\mathcal{N}\alpha/3)$. Indeed, for the square lattice, $C_0\approx 4.517$ and $C_{d=a}\approx -0.164$ so that $(C_0+C_a)/4\pi\approx 0.346$, very close to $1/3$.\\
In the bulk, coupling a layer with at least the two nearest neighboring ones, 
one could naively expect to have $(C_0+2C_a)/4\pi$, which is even closer to $1/3$, the 3D factor.

\section{The Fresnel coefficients of an insulating bilayer crystal}
\subsection{Free-standing bilayer crystal}
We want to express the Fresnel coefficients in terms of $\chi$. The best way to do this is to write eqs. (\ref{Bilayer_non-interacting}), (\ref{Bilayer_interacting}) and (\ref{honeycomb_interacting})  in term of $\chi$ and to solve them in this form.
\begin{eqnarray}
\label{Bilayer_Fresnel}
&E&_{i}+E_{r}=E_{+}+E_{-} \\ &E&_{i}+E_{r}=\frac{P_1}{\chi d \epsilon_{0}}+\frac{(P_1-P_2) C_F}{\epsilon_0} \nonumber \\
&H&_{i}-H_{r}=H_{+}-H_{-}+i\omega P_1  \nonumber \\
&E&_{+}e^{-ikd}+E_{-}e^{ikd}=E_{t};\nonumber \\ &E&_{t}=\frac{P_2}{\chi d \epsilon_{0}}-\frac{(P_1-P_2) C_F}{\epsilon_0} \nonumber \\
&H&_{+}e^{-ikd}-H_{-}e^{ikd}=H_{t}+i\omega P_2 \nonumber
\end{eqnarray}
Here $C_F$ has the dimension of the inverse of a distance. Even in the case of the honeycomb lattice (eqs. (\ref{honeycomb_interacting})), we have 6 equations instead of 8. For the non-interacting case, $C_F=0$ $\rm m^{-1}$. For the interacting case, for the square and the triangular lattices,
\begin{equation}
C_F=\frac{C_d}{4 \pi a^3 N}
\end{equation}
while the value for the honeycomb lattice is reported in Appendix II.
Defining $r_{s}=E_{r}/E_{i}$, $t_{s}=E_{t}/E_{i}$ as the reflection and the transmission coefficients, the non-interacting case appears as a natural extension of the monolayer. We obtain 
\begin{eqnarray}
\label{Fresnel_non_interacting_r}
&r&_s=\frac{r_1+r_2(t_1+r_1)e^{-2ikd}}{1-r_1 r_2e^{-2ikd}} \\
\label{Fresnel_non_interacting_t}
&t&_s=\frac{t_1t_2e^{-ikd}}{1-r_1 r_2e^{-2ikd}} 
\end{eqnarray}
where the subscripts $1$ ($2$) refers to the first (second) crystal plane met by the incident wave, and $r_1= r_2, t_1=t_2$ are respectively the reflection and tranmission coefficients for a free-standing single layer crystal (provided by formula (2) of ref \cite{Merano16}, where the surface susceptibility must be replaced with $\chi \cdot d$). For the interacting case we find
\begin{equation}
\label{Fresnel_interacting}
r_s=A+D; \quad t_s=B-D
\end{equation}
where $A$ and $B$ are respectively equal to (\ref{Fresnel_non_interacting_r}) and (\ref{Fresnel_non_interacting_t}), and $D$ is given by
\begin{eqnarray}
\label{D}
D&=&\frac{2kd(e^{ikd}-1)^2\chi^2}{e^{ikd}(kd\chi -2i)-kd\chi} \cdot \\ &\cdot&\frac{d C_{F}}{-ikd \chi +e^{ikd}(4C_{F}d\chi +ikd\chi +2)} \nonumber
\end{eqnarray}
As expected, due to the interaction, the Fresnel coefficients now depend explicitly also on $C_F$. This occurs only for terms at the order of $k^3d^3$ or bigger since their Taylor expansions 
\begin{eqnarray}
\label{Fresnel_Taylor}
&r&_s=-i\chi kd -\chi(1+\chi)k^2d^2 +O(k^3d^3) \\
&t&_s=1-i(1+\chi)kd-\frac{1}{2}(1+2\chi+2\chi^2)k^2d^2+O(k^3d^3) \nonumber
\end{eqnarray}
are the same for the non-interacting and the interacting case up to the second order, apart from the different expression of $\chi$ in the two cases.

\subsection{Bilayer crystal on a substrate}
\subsubsection{Semi-infinite substrate}
We consider now the case of a bilayer crystal deposited on a homogeneous transparent medium ($n_1=n_2$) which fills the half-space below it. As it was done in \cite{Blake2007, Blake2011} we assume that we can neglect the interaction of the 2D crystal with the substrate. With respect to eqs. (\ref{Bilayer_Fresnel}) only the relation between $\vec{\emph{\textbf{E}}}$ and $\vec{\emph{\textbf{H}}}$ in the transmitted waves changes
\begin{eqnarray}
\frac{\eta}{n_1}\vec{\textbf{\emph{H}}}_{t}=\hat{s}_{t}\wedge\vec{\textbf{\emph{E}}}_{t};
\end{eqnarray}

\subsubsection{Stratified substrate}
For comparison with the experimental data it is also useful to consider the case of a bilayer deposited on a stratified medium (fig. 1, $n_1\neq n_2$)
\begin{eqnarray}
\label{Bilayer_interacting_stratified}
&E&_{i}+E_{r}=E_{+}+E_{-} \\ &E&_{i}+E_{r}=\frac{P_1}{\chi d \epsilon_{0}}+\frac{(P_1-P_2) C_F}{\epsilon_0} \nonumber \\
&H&_{i}-H_{r}=H_{+}-H_{-}+i\omega P_1  \nonumber \\
&E&_{+}e^{-ikd}+E_{-}e^{ikd}=E_{1+}+E_{1-} \nonumber \\ &E&_{1+}+E_{1-}=\frac{P_2}{\chi d \epsilon_{0}}-\frac{(P_1-P_2) C_F}{\epsilon_0} \nonumber \\
&H&_{+}e^{-ikd}-H_{-}e^{ikd}=H_{1+}-H_{1-}+i\omega P_2 \nonumber \\
&E&_{1+}e^{-i\beta}+E_{1-}e^{i\beta}=E_{t}; \quad H_{1+}e^{-i\beta}-H_{1-}e^{i\beta}=H_{t} \nonumber
\end{eqnarray} 
where $\beta=kn_{1}h$ and $h$ is the thickness of medium 1. For the non interacting case, the Fresnel coefficients for these two types of substrates are still provided by (\ref{Fresnel_non_interacting_r}) and (\ref{Fresnel_non_interacting_t}). The only difference is that for the semi-infinite substrate, $r_2$ and $t_2$ must be replaced with formula (6) of \cite{Merano16} and for the stratified substrate with formulas obtained starting from the equation system (7) of \cite{Merano16}. Taylor's expansions (see Appendix II) of these expressions, for the non-interacting and the interacting case, are identical up to the second order in $kd$. Only the value of $\chi$ is different in the two cases. The first order terms of these expansions are the Fresnel coefficients of the substrate. 

\section{Analysis of optical contrast measurements}
In fig. (\ref{Structure}) the crystal structure of the bilayer hBN is reported. The dimensions of the unit cell are: $a$ = 0.25 nm and $c$ = 0.666 nm. The unit cell is bimolecular, with each atomic layer consisting of a flat network of $\rm B_3N_3$ hexagons with an interplanar distance of $d=c/2$ \cite{Pease52}. Figure (\ref{Exp}) shows variations of the optical contrast (for the definition of this quantity see \cite{Blake2007}) in the spectral range 410 nm $<\lambda <$ 740 nm for monolayer and bilayer hBN on top of a $\rm SiO_2/Si$ wafer with a nominal thickness of 290 nm. Dots are the experimental data that have been extracted from ref. \cite{Blake2011} via software digitization. The same paper reports the theoretical fits to these experimental data based on a slab model, and it assumes the same refractive index for the monolayer and the bilayer crystals. In practice ref. \cite{Blake2011} assumes that the equality holds in eq. (\ref{chivschis}).

\begin{figure}
\includegraphics{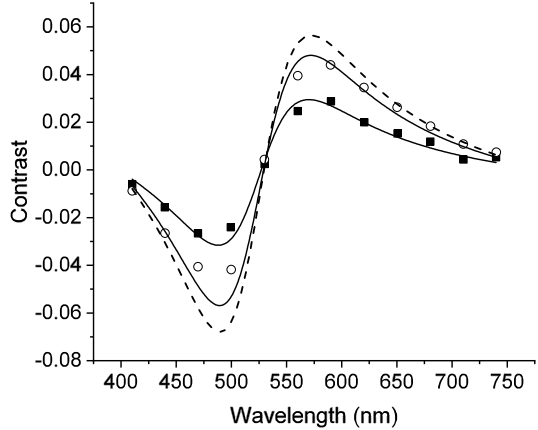}
\caption{Optical contrast of monolayer and bilayer hBN on top of a $\rm SiO_2/Si$ wafer. Solid dots: experimental data for the monolayer \cite{Blake2011}; open dots: experimental data for the bilayer \cite{Blake2011}. Solid lines: best theoretical fits assuming, respectively, $\chi_s = \textrm{1.3 nm}$ for the monolayer and $\chi \, d = \textrm{1.1 nm}$ for the bilayer (see the text). Dashed line: theoretical fit assuming no variation of the susceptibility from the monolayer to the bilayer}
\label{Exp2}
\end{figure}

The value of $\chi_s$ for the monolayer has already been deduced in ref. \cite{Merano316}. The solid line is the best theoretical fit assuming $\chi_s = 1.3\cdot 10^{-9}$ m. The value of the surface conductivity was extimated to be $\sigma \leq 2\cdot 10^{-6}\ \Omega^{-1}$, confirming that we are dealing with an insulating dielectric material. Starting from the Fresnel coefficients derived from eqs. (\ref{Bilayer_interacting_stratified}), the best theoretical fit (solid line) for the bilayer gives $\chi$ = 3.34, so that we have
\begin{equation}
\label{chivschis_exp}
\chi_s = \textrm{1.3 nm} >\chi \, d = \textrm{1.1 nm}
\end{equation}
If we assume no variation of the susceptibility from the monolayer to the bilayer (i.e. $\chi \, d = \textrm{1.3 nm}$), the theoretitical fit that we obtain is the dashed line in fig. \ref{Exp2}. The experimental data are clearly consistent with a variation of the susceptibility from the monolayer to the bilayer and more specifically with our eq. (\ref{chivschis}). The optical contrast measurements are very sensible to the $\rm SiO_2$ thickness, as discussed in Appendix IV.

\subsection{Atomic polarizabilities}
Having both the $\chi_s$ from the optical contrast measurements of the monolayer \cite{Blake2011, Merano316} and the $\chi$ for the bilayer, we can try to deduce the atomic polarizabilities. From the equations
\begin{equation}
\label{system}
\chi_s=\frac{N\Big(\alpha_1+\alpha_2-\frac{\alpha_1\alpha_2\left(C_0^{(1)}-C_0^{(2)}\right)}{2\pi a^3}\Big)}{1-\frac{C_0^{(1)}(\alpha_1+\alpha_2)}{4\pi a^3}+\frac{\alpha_1\alpha_2\left(C_0^{(1)2}-C_0^{(2)2}\right)}{(4\pi a^3)^2}}
\end{equation}
\begin{equation}
\chi=\frac{N\Big(\frac{\alpha_1+\alpha_2}{d}-\frac{\alpha_1\alpha_2\left(C_0^{(1)}-C_0^{(2)}-C_d^{(1)}+C_d^{(2)}\right)}{2\pi a^3d}\Big)}{1-\frac{(C_0^{(1)}+C_d^{(2)})(\alpha_1+\alpha_2)}{4\pi a^3}+\frac{\alpha_1\alpha_2\left((C_0^{(1)}+C_d^{(2)})^2-(C_0^{(2)}+C_d^{(1)})^2\right)}{(4\pi a^3)^2}}\nonumber
\end{equation}
we can now extract the values of $\alpha_1$ and $\alpha_2$ that are the only unknown parameters of these two equations. We obtain $\alpha_1=\rm 1.1 \cdot 10^{-24} cm^{-3}$ and $\alpha_2 = \rm 19.5 \cdot 10^{-24} cm^{-3}$, which have the right order of magnitude if compared with the static calculations reported in ref. \cite{Dalgarno62}. Of course, as for the 3D case, atomic polarizabilities require a full quantum dynamical approach for a proper treatment.      

\section{Conclusions}  
In this paper, we provide a complete classical description of the optical response of a bilayer crystal. We compute both the local and the macroscopic fields. We find that, if the distance $d$ between the two planes far exceeds the lattice constant $a$, they can be macroscopically treated as two separated monolayers. In this case, for both layers, the local field is connected to the macroscopic field via an expression identical to the one for the monolayer. On the other hand, for shorter distances, although it is still possible to define a surface polarization and hence a surface current for each single plane, the local fields are coupled to the macroscopic fields in both layers. As a consequence of this, it is meaningful to provide a volume susceptibility rather than a surface susceptibility. Remarkably, the expression for the volume susceptibility is still very simple, and it depends on the interaction parameter $C_d$ (or $C^{(1)}_d$ and $C^{(2)}_d$), which rapidly decreases with $d$, exhibiting an evanescent-wave character. Interestingly, for the Fresnel coefficients the effect of the coupling between the layers in the long wavelength limit (small $k$) is fully contained in the susceptibility. Even if we use a classical approach to compute the susceptibility, the expression of the Fresnel coefficients that we derive has a general validity because they concern the macroscopic fields.

A comparison of our theory with the optical contrast measurements \cite{Blake2011} confirms that the interaction between the two atomic planes alters the optical response of the bilayer crystal. Its susceptibility is indeed modified from that expected for non-interacting layers by an experimentally appreciable amount.

The approach reported here is valid also for all-dielectric, double-layer metasurfaces \cite{Holloway12} and for bilayer heterostructures \cite{Geim14}, and we believe that, in these contexts, our work can lead to some potential applications. Some questions still remain about how two layers of different materials interact. Also twisted bilayers \cite{Ahn2018} have macroscopic properties that are different from each other. Because metasurfaces can be designed to have total reflection for an incident wave, it is possible to trap and guide electromagnetic energy in a region between two metasurfaces. Monolayer crystals can support in general both transverse electric and transverse magnetic modes. Bilayer crystals, twisted bilayer crystals, or bilayer heterostructures are, therefore, promising devices for designing new ultra-thin waveguides.

Based on our findings, it would be interesting to compare experimentally the optical contrast of a BN bilayer with that of two BN monolayers separated by a distance much larger than $a$, for instance in a system composed sequentially by a substrate, a monolayer, a dielectric film, and a monolayer. In the latter case, we expect that the optical response is well described by the surface susceptibility of a single monolayer. The interaction between the layers and its dependence on the distance, Eq. \ref{Cd}, might be addressed experimentally also using bilayer heterostructures.

We think that the same approach developed in this paper can be extended to study multilayer structures up to a thickness where the bulk susceptibility is found. Some questions anyway are still to be solved. All the fields that we have considered in this paper travel in the vacuum at speed $c$, so one may wonder how a polarization wave that propagates with velocity $c/n$ emerges ($n$ being the refractive index), and how many layers we need in order to have a layer independent refractive index. For a bilayer crystal we do not need to consider the speed of the polarization wave because the macroscopic field is simply given by the incident plus the reflected fields in the first layer and by the transmitted field in the second layer, but, in the case of many layers, we would need an expression for the macroscopic field inside the crystal. 

\section{ACKNOWLEDGMENTS}
L.D. acknowledges financial support from Dipartimento di Fisica e Astronomia G. Galilei, Universit$\rm \grave{a}$ Degli Studi di Padova, funding BIRD2016, M. M. acknowledges financial support from Dipartimento di Fisica e Astronomia G. Galilei, Universit$\rm \grave{a}$ Degli Studi di Padova, funding BIRD170839/17 and from MIUR, funding FFARB.

\section{Author contributions}
MM conceived the idea. Both authors discussed the physical model and wrote the paper.

\section{Appendix I}
\subsection{Macroscopic Theory: honeycomb lattice}
\begin{eqnarray}
\label{honeycomb_interacting}
&E&_{i}+E_{r}=E_{+}+E_{-} \\ &E&_{i}+E_{r}=E^{(1, 1)}_{loc}-\nonumber \\ &-& \frac{\alpha_1(E^{(1, 1)}_{loc}C^{(1)}_{0}+E^{(2, 1)}_{loc}C^{(2)}_{d})+\alpha_2(E^{(1, 2)}_{loc}C^{(2)}_{0}+E^{(2, 2)}_{loc}C^{(1)}_{d})}{4\pi a^3}  \nonumber \\
&E&_{i}+E_{r}=E^{(1, 2)}_{loc}-\nonumber \\ &-& \frac{\alpha_1(E^{(1, 1)}_{loc}C^{(2)}_{0}+E^{(2, 1)}_{loc}C^{(1)}_{d})+\alpha_2(E^{(1, 2)}_{loc}C^{(1)}_{0}+E^{(2, 2)}_{loc}C^{(2)}_{d})}{4\pi a^3}  \nonumber \\
&H&_{i}-H_{r}=H_{+}-H_{-}+i\frac{k}{\eta}N(\alpha_1 E^{(1, 1)}_{loc}+\alpha_2 E^{(1, 2)}_{loc})  \nonumber \\
&E&_{+}e^{-ikd}+E_{-}e^{ikd}=E_{t};\nonumber \\ &E&_{t}=E^{(2, 1)}_{loc}-\nonumber \\ &-& \frac{\alpha_1(E^{(1, 1)}_{loc}C^{(2)}_{d}+E^{(2, 1)}_{loc}C^{(1)}_{0})+\alpha_2(E^{(1, 2)}_{loc}C^{(1)}_{d}+E^{(2, 2)}_{loc}C^{(2)}_{0})}{4\pi a^3}  \nonumber \\
&E&_{t}=E^{(2, 2)}_{loc}-\nonumber \\ &-& \frac{\alpha_1(E^{(1, 1)}_{loc}C^{(1)}_{d}+E^{(2, 1)}_{loc}C^{(2)}_{0})+\alpha_2(E^{(1, 2)}_{loc}C^{(2)}_{d}+E^{(2, 2)}_{loc}C^{(1)}_{0})}{4\pi a^3}  \nonumber \\
&H&_{+}e^{-ikd}-H_{-}e^{ikd}=H_{t}+i\frac{k}{\eta}N(\alpha_1 E^{(2, 1)}_{loc}+\alpha_2 E^{(2, 2)}_{loc}) \nonumber
\end{eqnarray}
The non-interacting case corresponds to $C^{(1)}_d=C^{(2)}_d=$ 0. 

\section{Appendix II}
\subsection{Expression of $C_F$ for the honeycomb lattice}
Calling 
\begin{eqnarray}
\Delta=C^{(1)}_0-C^{(2)}_0\\ 
\Delta_d=C^{(1)}_d-C^{(2)}_d
\end{eqnarray}
we have
\begin{widetext}
\begin{eqnarray}
\label{hb_xi}
C_F=\frac{C_{d}^{(2)}}{4\pi a^3 N}  
+\frac{\Delta_d  \alpha_1 \alpha_2 \left(16 \pi^2 a^6
+\left(\Delta^2-\Delta_d^2\right) \alpha_1 \alpha_2-4 \pi a^3 \Delta  
(\alpha_1+\alpha_2)\right)}{8 N \pi a^3 ((\Delta_d-\Delta) 
\alpha_1 \alpha_2+2 \pi a^3  (\alpha_1+\alpha_2)) (2 \pi a^3 (\alpha_1+\alpha_2)
-(\Delta+\Delta_d) \alpha_1 \alpha_2)}
\end{eqnarray}
\end{widetext}
Notice that, for $\alpha_1=\alpha_2$, this equation reduces to 
$C_F=\frac{C_d^1+C_d^2}{8\pi a^3 N}$ 
so that the equations for the macroscopic electric fields are simply 
\begin{eqnarray}
\label{hb_xi_eq}
&E&_{i}+E_{r}=\frac{P_1}{\chi d \epsilon_{0}}
+\frac{(P_1-P_2) (C_d^{(1)}+C_{d}^{(2)})}{8\pi a^3 N \epsilon_0}
\\
&E&_{t}=\frac{P_2}{\chi d \epsilon_{0}}
-\frac{(P_1-P_2) (C_d^{(1)}+C_{d}^{(2)})}{8\pi a^3 N \epsilon_0}
\end{eqnarray}

\section{Appendix III}
\subsection{Taylor expansion of the Fresnel coefficients: semi-infinite substrate}
\begin{eqnarray}
\label{Fresnel_Taylor_substrate}
&r_s&=-\frac{n_1-1}{n_1+1}+\frac{2i(n_1^2-1-2\chi)}{(n_1+1)^2}kd+ \\&+&\frac{2(n_1^2-1-2\chi)(n_1+1+2\chi)}{(n_1+1)^3}k^2d^2+O(k^3d^3) \nonumber \\
&t_s&=\frac{2}{n_1+1}-\frac{2i(n_1+1+2\chi)}{(n_1+1)^2}kd \nonumber+ \\&-&\frac{1+2n_1+n_1^2+6\chi+4n_1\chi-2n^2\chi+8\chi^2}{(n_1+1)^3}k^2d^2+ O(k^3d^3) \nonumber
\end{eqnarray} 
The first terms of the expansions are the Fresnel coefficients of the substrate. It is easy to verify that the same holds for a stratified substrate.

\section{Appendix IV}
\subsection{Optical contrast as a function of the $\rm SiO_2$ thickness}
\begin{figure}
\includegraphics{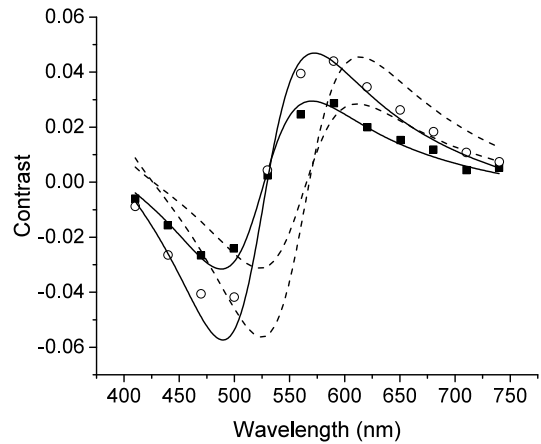}
\caption{Optical contrast of mono and bilayer hBN on top of a $\rm SiO_2/Si$ wafer. Solid dots: experimental data for the monolayer \cite{Blake2011}, open dots experimental data for the bilayer \cite{Blake2011}. Dash lines: best theoretical fits for a $\rm SiO_2$ thickness of 290 nm. Solid lines: best theoretical fits for a $\rm SiO_2$ thickness of 270 nm.}
\label{Exp}
\end{figure}

The dashed line in Fig. \ref{Exp} is the best theoretical fit, for the optical contrast data of a monolayer, assuming $\chi_s = 1.3\cdot 10^{-9}$ m and the nominal $\rm  SiO_{2}$ thickness of 290 nm. The only way to improve the fit is by varying the $\rm SiO_{2}$ thickness, showing that the spectral position of the optical contrast curve depends much on the substrate. The solid line is the theoretical fit for the same values of $\chi_s$ and $\sigma$ but a $\rm SiO_{2}$ thickness of 270 nm. Indeed, we noticed that by increasing the thickness of the substrate, the optical contrast curve translates towards the infrared, and new zeros (or new oscillations as a function of the wavelength) appear on the blue side. Starting from the Fresnel coefficients deduced from eqs. (\ref{Bilayer_interacting_stratified}), the best theoretical fit (solid line) for the bilayer provides a $\chi$ = 3.34 and a $\rm  SiO_{2}$ thickness of 270 nm. For the sake of completeness, the theoretical fit for the nominal thickness of 290 nm is shown as a dashed line. 
\bibliography{letter}
\end{document}